\documentclass[letterpaper,11pt]{article}
\usepackage[margin=1in]{geometry}
\usepackage[utf8]{inputenc}

\usepackage[affil-it]{authblk}
\usepackage{amsmath,amsfonts,amssymb,amsthm}
\usepackage{mathtools}
\usepackage{bbm}
\usepackage[dvipsnames]{xcolor}
\usepackage[colorlinks=true,linkcolor=blue,urlcolor=NavyBlue,citecolor=Fuchsia]{hyperref}
\usepackage{enumitem}
\usepackage{empheq}
\newcommand*\widefbox[1]{\fbox{\hspace{2em}#1\hspace{2em}}}

\usepackage{tikz}
\usetikzlibrary{positioning}
\usetikzlibrary{shapes}
\usetikzlibrary{calc}
\usetikzlibrary{cd}
\usetikzlibrary{decorations.markings}

\usepackage{tocloft}

\usepackage{bm}
\usepackage{braket}
\usepackage{tikz}
\usepackage{algorithm}
\usepackage{algpseudocode}

\usepackage{sectsty}
\sectionfont{\large}

\let\originalleft\left
\let\originalright\right
\renewcommand{\left}{\mathopen{}\mathclose\bgroup\originalleft}
\renewcommand{\right}{\aftergroup\egroup\originalright}

\newcommand{\E}{\mathbb{E}}
\newcommand{\smean}{\mathrm{SMean}}

\newcommand{\var}{\mathrm{Var}}
\newcommand{\svar}{\mathrm{SVar}}
\newcommand{\cov}{\mathrm{Cov}}

  \usepackage{float}
  \usepackage{afterpage}
  \usepackage{array}
  \usepackage{longtable}
  \usepackage{booktabs}
  \usepackage{multirow}
  \usepackage{tabularx}
  
  \newcolumntype{P}[1]{>{\centering\arraybackslash}p{#1}} 
  \newcolumntype{M}[1]{>{\centering\arraybackslash}m{#1}} 
  \newcolumntype{B}[1]{>{\centering\arraybackslash}b{#1}} 
  
  \usepackage{dcolumn}
  \newcolumntype{.}{D{.}{.}{-1}} 

\newcommand{\fref}[1]{\hyperref[#1]{Figure~\ref*{#1}}}
\newcommand{\Fref}[1]{\hyperref[#1]{Figure~\ref*{#1}}}
\newcommand{\sref}[1]{\hyperref[#1]{Section~\ref*{#1}}}
\newcommand{\Sref}[1]{\hyperref[#1]{Section~\ref*{#1}}}
\newcommand{\tref}[1]{\hyperref[#1]{Table~\ref*{#1}}}
\newcommand{\Tref}[1]{\hyperref[#1]{Table~\ref*{#1}}}
\newcommand{\aref}[1]{\hyperref[#1]{Appendix~\ref*{#1}}}
\newcommand{\Aref}[1]{\hyperref[#1]{Appendix~\ref*{#1}}}
\newcommand{\thref}[1]{\hyperref[#1]{Theorem~\ref*{#1}}}
\newcommand{\Thref}[1]{\hyperref[#1]{Theorem~\ref*{#1}}}
\newcommand{\alref}[1]{\hyperref[#1]{Algorithm~\ref*{#1}}}
\newcommand{\Alref}[1]{\hyperref[#1]{Algorithm~\ref*{#1}}}
\newcommand{\defref}[1]{\hyperref[#1]{Definition~\ref*{#1}}}

\def\reportnumber{FERMILAB-FN-1273-ETD}

\newcommand{\reportnumberstring}{}
\ifx\reportnumber\undefined\else
  \ifx\reportnumber\empty\else
    \renewcommand{\reportnumberstring}{{%
      \sffamily%
      {\small Report No.:} %
      {\footnotesize\textbf{\reportnumber}}%
    }}
  \fi
\fi

\usepackage[
  angle=0,scale=1,color=black,opacity=1,%
  firstpage=true,
  contents=
]{background}

\SetBgPosition{current page.north}
\SetBgVshift{-0.45in}

\SetBgContents{%
  \color{black!60}%
  \begin{minipage}{1.1\textwidth}%
    \hfill\reportnumberstring
  \end{minipage}%
}

\let\oldtitle\title
\let\oldaffil\affil
\let\olddate\date
\renewcommand{\title}[1]{\oldtitle{\vspace{-3em}\large\textsc{#1}\vspace{-.2em}}}
\renewcommand{\date}[1]{\olddate{\vspace{-.5em}#1}}
\renewcommand{\affil}[1]{\oldaffil{{\normalfont\small\vspace{-1em}#1}}}

\title{Comment on ``An implementation of neural simulation-based
inference\\for parameter estimation in ATLAS''}
\author{Prasanth Shyamsundar}
\affil{Fermi National Accelerator Laboratory, Batavia, IL 60510, USA}
\date{{\small May 25, 2025}}

\begin{document}

\maketitle
\begin{abstract}
 The paper titled ``An implementation of neural simulation-based inference for parameter estimation in ATLAS'' by the ATLAS collaboration (\href{https://arxiv.org/abs/2412.01600v1}{arXiv:2412.01600v1 [hep-ex]}) describes the implementation of neural simulation-based inference for a measurement analysis performed by ATLAS. The uncertainties in the analysis arising from the finiteness of the simulated datasets are estimated using a novel double-bootstrapping technique described in that work. In the present comment, it is claimed and demonstrated, using a toy example, that the double-bootstrapping technique does not actually capture the aforementioned uncertainties.
\end{abstract}

\section{Introduction}

Ref.~\cite{ATLAS:2024jry} presents a measurement of off-shell Higgs boson production in the $H^* \rightarrow ZZ \rightarrow 4\ell$ decay channel, by the ATLAS collaboration, using the neural
simulation-based inference (NSBI) technique. This is among the first public results resulting from the application of NSBI in experimental collider physics. The technical details of the NSBI implementation are described in Ref.~\cite{ATLAS:2024ynn}.
The uncertainties arising from the finiteness of the simulated dataset used in the analysis, often referred to as mc-stat uncertainties (i.e., Monte Carlo statistical uncertainties), are estimated using a novel double-bootstrapping procedure described in Ref.~\cite{ATLAS:2024ynn}. The present comment identifies a technical flaw, which leads to the mc-stat uncertainties being completely missed by the double-bootstrapping approach.

\section{Description of the two-level bootstrapping approach of Ref.~\cite{ATLAS:2024ynn}}\label{sec:desc_of_orig_procedure}
Let $D$ be a given simulated training dataset of size $N$, containing possibly weighted, labeled background and signal events. The background and signal distributions are chosen to not have any unknown parameters. The tasks under consideration are to
\begin{enumerate}[label=\alph*)]
 \item Estimate the likelihood ratio function $r(x)$ given by
 \begin{align}
  r(x)\equiv \frac{p_S(x)}{p_B(x)}\,,
 \end{align}
 using the available simulated dataset $D$, by training one or more neural networks. Here $x$ represents the input event-attributes, and $p_B$ and $p_S$ are the background and signal probability density functions of $x$.
 \item Perform an analysis using the estimate for $r$, to get an estimate for a theory parameter $\mu$, which corresponds to the signal strength.
 \item Estimate uncertainties in the estimates of $r$ and $\mu$ arising from the finiteness of the training dataset and random fluctuations in the training of the neural networks. 
\end{enumerate}
\paragraph{Estimating $r$ and $\mu$.}
The simulated dataset $D$ is bootstrapped independently $M$ times to get datasets $D_{(\text{boot},1)}, \dots, D_{(\text{boot},M)}$ of sizes close to $N$ (within possible Poisson fluctuations). For each $m\in\{1,\dots,M\}$, the dataset $D_{(\text{boot},m)}$ is used to train a neural network estimator $\hat{r}_{(\text{boot},m)}$ for the function $r$. The details of the neural network training are irrelevant for the purposes of this comment. The subscript ``$\text{boot}$'' in $\hat{r}_{(\text{boot},m)}$ indicates that it was trained on a bootstrapped dataset. The different neural networks are then averaged to get a merged estimate for $r$ given by\footnote{The arguments in the present comment also work for some alternative ways to merge the $\hat{r}_{(\text{boot},m)}$-s, e.g.,
\begin{align*}
 \hat{r}_\text{boot-geom-avg}(x) &\equiv \left[\prod_{m=1}^M \hat{r}_{(\text{boot},m)}(x)\right]^{1/M} = \exp\left(\frac{1}{M}\,\sum_{m=1}^M \ln\left(\Big.\hat{r}_{(\text{boot},m)}(x)\right)\right)\,.
\end{align*}}
\begin{align}
 \hat{r}_\text{boot-avg}(x) &\equiv \frac{1}{M}\,\sum_{m=1}^M \hat{r}_{(\text{boot},m)}(x)\,. \label{eq:r_boot_avg}
\end{align}
Let $\hat{r}_\text{boot-avg}$ be the final estimate for $\hat{r}$.
This estimated likelihood ratio function $\hat{r}_\text{boot-avg}$ is used to get an estimate $\hat{\mu}_\text{boot-avg}$ for the parameter $\mu$. The details of this step
are irrelevant for the purposes of this comment. The subscipt ``$\text{boot-avg}$'' in $\hat{\mu}_\text{boot-avg}$ is used to denote that it was computed using $\hat{r}_\text{boot-avg}$. The map from $\hat{r}_\text{boot-avg}$ to $\hat{\mu}_\text{boot-avg}$ could possibly be non-deterministic.

\paragraph{Estimating the uncertainties.}

The uncertainties in $\hat{r}_\text{boot-avg}(x)$ and $\hat{\mu}_\text{boot-avg}$ are estimated using a second round of bootstrapping as follows. Let $E_\text{boot}$ represent the ensemble of neural-network-based likelihood ratio estimates:
\begin{align}
 E_\text{boot} &\equiv \left(\Big.\hat{r}_{(\text{boot},1)},\dots,\hat{r}_{(\text{boot},M)}\right)\,.
\end{align}
$E_\text{boot}$ is bootstrapped independently $K$ times to get $K$ different double-bootstrapped ensembles:
\begin{align}
 E_{\text{boot-boot},k} &\equiv \left(\Big.\hat{r}_{(\text{boot-boot},k,1)},\dots,\hat{r}_{(\text{boot-boot},k,M)}\right)\,,\qquad\qquad\forall k\in\{1,\dots,K\}\,.
\end{align}
Importantly, each $\hat{r}_{(\text{boot-boot},k,m)}$ is simply sampled from the ensemble $E_\text{boot}$, with replacement. In other words, for any $k\in\{1,\dots,K\}$ and $m\in\{1,\dots,M\}$, there exists an $m'\in\{1,\dots,M\}$, such that $\hat{r}_{(\text{boot-boot},k,m)}\equiv \hat{r}_{(\text{boot},m')}$. Each double-bootstrapped ensemble $E_{\text{boot-boot},k}$ is then used to estimate a merged likelihood ratio estimator as follows:
\begin{align}
 \hat{r}_{(\text{boot-boot-avg},k)}(x) &\equiv \frac{1}{M}\,\sum_{m=1}^M \hat{r}_{(\text{boot-boot},k,m)}(x)\,.
\end{align}
For any given $x$, the sample standard deviation of the $\hat{r}_{(\text{boot-boot-avg},k)}(x)$-s is used as the uncertainty estimate for $\hat{r}_\text{boot-avg}(x)$.

The same $\mu$-estimation procedure from before (which was used to get $\hat{\mu}_\text{boot-avg}$ from $\hat{r}_\text{boot-avg}$) is performed using each $\hat{r}_{(\text{boot-boot-avg},k)}$ to get $K$ different $\mu$ estimates, denoted as $\hat{\mu}_{(\text{boot-boot-avg},k)}$ for $k\in\{1,\dots,K\}$. The sample standard deviation of the $\hat{\mu}_{(\text{boot-boot-avg},k)}$-s is used as an uncertainty estimate for $\hat{\mu}_{\text{boot-avg}}$.

The uncertainty estimate for $\hat{\mu}_{\text{boot-avg}}$ (at different possible true values of $\mu$) is introduced as a systematic uncertainty in the final analysis using the spurious signal approach \cite{ATLAS:2020ocz}; the details of this step are irrelevant for the purposes of this comment.

\section{Claim of the present comment}\label{sec:main_arguments}

The main claim of this comment is as follows: The uncertainty estimation procedure described above does not capture \textbf{any} of the mc-stat uncertainties in $\hat{r}_\text{boot-avg}$ and $\hat{\mu}_\text{boot-avg}$.

This claim may seem counterintuitive, considering that bootstrapping is a technique that specializes in capturing statistical uncertainties. However, one can motivate the claim as follows. Consider the limit where $N$ (the number of events in the original training dataset) is kept fixed, and $M$ (the number of neural networks trained using bootstrapped data) is increased indefinitely. In this limit, for any given $x$, the sample standard deviation of the $\hat{r}_{(\text{boot-boot-avg},k)}(x)$-s tends to $0$. However, since mc-stat is related to the finiteness of $N$, the size of mc-stat should have a non-zero lower bound that only decreases as $N$ is increased. This is a sign that mc-stat is not estimated correctly by the double-bootstrapping technique.

To understand why the technique does not work as intended, it may be instructive to review the assumptions behind the standard bootstrapping approach for quantifying uncertainties.

\paragraph{The tangible and intangible assumptions behind bootstrapping.} Abstractly, bootstrapping works as follows. Based on a statistical experiment probing a probability distribution $p_\mathrm{true}$\footnote{$p_\mathrm{true}$ is the distribution of the labeled training datapoints for the first bootstrapping, and the distribution of trained neural networks for the second bootstrapping.}, one estimates a property of $p_\mathrm{true}$, say $T[p_\mathrm{true}]$ as $\hat{T}$. Here $T$ could be a 1-dimensional or a multidimensional parameter, a function, etc. In order to estimate the statistical properties (e.g., variance) of $\hat{T}$, one would ideally want to repeat the experiment independently many times and get a sample of $\hat{T}$-s. If performing such repetitions with $p_\mathrm{true}$ is not feasible, then one could create a surrogate distribution $p_\mathrm{surrogate}$, which, in vague terms, is similar to $p_\mathrm{true}$. Subsequently, one can perform multiple repetitions of the experiment on $p_\mathrm{surrogate}$ to get a sufficiently large sample of $\hat{T}$-s corresponding to $p_\mathrm{surrogate}$. \emph{Certain} statistical properties of $\hat{T}$ under $p_\mathrm{true}$ are (a) assumed to be sufficiently close to their corresponding values under $p_\mathrm{surrogate}$ and hence (b) estimated from the surrogate sample of $\hat{T}$-s.

In the standard bootstrapping approach of resampling with replacement, infinite copies of the original dataset implicitly defines the surrogate $p_\mathrm{surrogate}$.\footnote{Training a generative machine learning model could be yet another way to construct a surrogate model for bootstrapping purposes.} Sampling from this surrogate can also be performed by applying Poisson perturbations to the event weights in the original dataset; this is the approach used in Refs.~\cite{ATLAS:2024jry,ATLAS:2024ynn}. Some tangible assumptions
of bootstrapping are as follows:
\begin{enumerate}[label=\Roman*)]
 \item Each pseudo experiment (including the data sampling and the subsequent analysis) performed on the surrogate to compute a $\hat{T}$ is performed \textbf{exactly} like the original experiment, with the only difference being that the pseudo experiments probe $p_\mathrm{surrogate}$ instead of $p_\mathrm{true}$.
 \item The pseudo experiments performed on the surrogate distribution are independent of each other, conditional on $p_\mathrm{surrrogate}$ (i.e., after constructing and fixing $p_\mathrm{surrogate}$). This assumption is required since it is difficult to estimate statistical properties from dependent samples of $\hat{T}$.
\end{enumerate}
The first assumption is important for the purposes of the present comment. Some examples of requirements that could be imposed by the first assumption are as follows. If the original experiment sampled independent and identically distributed (iid) datapoints from $p_\mathrm{true}$, then each pseudo experiment must sample iid datapoints from $p_\mathrm{surrogate}$. Conversely, if the original experiment sampled \emph{dependent} datapoints from $p_\mathrm{true}$, in some specific manner, then the pseudo experiments must also sample dependent datapoints from $p_\mathrm{surrogate}$, in the same manner.\footnote{In this case, the construction of a $p_\mathrm{surrogate}$ similar to $p_\mathrm{true}$ could itself be tricky, if one only has access to dependent datapoints sampled from $p_\mathrm{true}$.} If the original experiment itself involved training multiple neural networks, then each pseudo experiment must also involve the same.

The intangible assumption in bootstrapping is that the chosen surrogate distribution $p_\mathrm{surrogate}$ is sufficiently similar to $p_\mathrm{true}$, in order to estimate the statistical properties of interest sufficiently accurately; this assumption could break down in some situations.
The present comment only focuses on violations of the tangible assumptions listed above.

\paragraph{Discussion focusing on the second level of bootstrapping.} Recall that the second bootstrapping is performed by simply resampling from $E_\text{boot}$ with replacement. In other words, the elements of each $E_{(\text{boot-boot},k)}$ are \textbf{independent} and identically sampled from the surrogate distribution implicitly defined by $E_\text{boot}$:
\begin{align}
 p\left(\hat{r}_{(\text{boot-boot},k,1)}, \dots, \hat{r}_{(\text{boot-boot},k,M)}~~\Big|~~E_\text{boot}\right) \equiv \prod_{m=1}^M p\left(\hat{r}_{(\text{boot-boot},k,m)}~~\Big|~~E_\text{boot}\right)\,.\label{eq:bootboot-iid}
\end{align}
On the other hand, the elements of $E_\text{boot}$, namely the $\hat{r}_{(\text{boot},m)}$-s, are \textbf{not independent} of each other:
\begin{align}
 p\left(\Big.\hat{r}_{(\text{boot},1)}, \dots, \hat{r}_{(\text{boot},M)}\Big.\right) &\not\equiv  \prod_{m=1}^M p\left(\hat{r}_{(\text{boot},m)}\Big.\right)\,.\label{eq:boot-notiid}
\end{align}
They are dependent on each other by virtue of being trained on datasets bootstrapped from the same original training dataset $D$. Equations \eqref{eq:bootboot-iid} and \eqref{eq:boot-notiid} indicate that assumption (I) of bootstrapping is violated for the purposes of estimating the overall uncertainties in $\hat{r}_\text{boot-avg}$ and $\hat{\mu}_\text{boot-avg}$, under the random procedure considered here for producing them.

The $\hat{r}_{(\text{boot},m)}$-s are in fact mutually independent of each other conditional on $D$, i.e., assuming that $D$ is fixed:
\begin{align}
 p\left(\Big.\hat{r}_{(\text{boot},1)}, \dots, \hat{r}_{(\text{boot},M)}~~\Big|~~D\right) &\equiv  \prod_{m=1}^M p\left(\hat{r}_{(\text{boot},m)}~~\Big|~~D\right)\,.\label{eq:boot-iid}
\end{align}
Equations \eqref{eq:bootboot-iid} and \eqref{eq:boot-iid} indicate that assumption (I) of bootstrapping is satisfied for the purposes of estimating the uncertainties \emph{conditional on} $D$. In other words, the double-bootstrapped uncertainties can capture the impact of various sources of randomness in the training procedure for a given $D$ (e.g., in the bootstrapping procedures, initialization of the networks' weights), but do not account for the statistical fluctuations in $D$. This leads to the claim that they do not cover any of the mc-stat uncertainties.


\paragraph{Discussion focusing on the first level of bootstrapping.} An argument that only focuses on the second bootstrapping may be unconvincing to the reader, since the first bootstrapping is where one resamples from $D$---this is typically the step that mimics and captures the effects of the statistical fluctuations in $D$. To analyze the first bootstrapping, consider the procedure used to compute $\hat{r}_\text{boot-avg}$ and $\hat{\mu}_\text{boot-avg}$ from $D$. This procedure involves (a) resampling $D$ to get $M$ bootstrapped datasets, (b) training $M$ neural networks, and (c) averaging the neural networks and performing subsequent analyses. Since this procedure is not independently repeated in each pseudo experiment performed with a $D_{(\text{boot},m)}$, one should not expect the resulting uncertainty estimates to capture mc-stat, again due to the violation of assumption (I) of bootstrapping.

\section{Demonstration using a toy example}
The discussions above may be generally unconvincing to the reader. It may also seem like (a) the present comment is being excessively pedantic about the theory of bootstrapping, or (b) the claim that the double-bootstrapping approach does not capture \textbf{any} of the mc-stat uncertainties is exaggerated. To address such skepticism, a toy example will be used to demonstrate the claims of the present comment.

Let $D\equiv (X_1,\dots, X_N)$, where $X_i$-s are iid datapoints sampled from the 1-dimensional normal distribution with mean $\theta$ and standard deviation $\sigma_X$. The goal is to estimate $\theta$ from $D$, and also estimate the corresponding uncertainty. $\theta$ is analogous to $r$ and $\mu$ in \sref{sec:desc_of_orig_procedure}. This is a simplified setup where (a) one only estimates a real-valued quantity instead of a function like $r$, and (b) the additional post-processing to get a $\mu$-estimate from an $r$-estimate is avoided. The normal distribution is analogous to the distribution of training datapoints in \sref{sec:desc_of_orig_procedure}, with the $X_i$-s being analogous to the labeled training datapoints. The dataset $D$ is resampled with replacement $M$ times to get
\begin{align}
 D_{(\text{boot},m)} &\equiv \left(X_{(\text{boot},m,1)},\dots, X_{(\text{boot},m,N)}\right)\,,\qquad\qquad\forall m\in{1,\dots,M}\,.
\end{align}
Next, each $D_{(\text{boot},m)}$ is used to estimate $\theta$ as follows:
\begin{align}
 \hat{\theta}_{(\text{boot},m)} &\equiv \left(\frac{1}{N}\sum_{i=1}^N X_{(\text{boot},m,i)}\right) + \epsilon_m\,,\label{eq:theta_boot}\\
 E_\text{boot} &\equiv \left(\Big.\hat{\theta}_{(\text{boot},1)}, \dots,\hat{\theta}_{(\text{boot},M)}\right)\,,
\end{align}
where $\epsilon_m$-s are independent normal random variables with mean $0$ and standard deviation $\sigma_\epsilon$. The $\hat{\theta}_{(\text{boot},m)}$-s are analogous to the neural-network-based $\hat{r}_{(\text{boot},m)}$-s in \sref{sec:desc_of_orig_procedure}. The $\epsilon_m$-s are analogous to random training errors, arising from various sources of randomness in the neural network training procedure. A merged estimate for $\theta$ is computed as
\begin{align}
 \hat{\theta}_\text{boot-avg} \equiv \frac{1}{M}\sum_{m=1}^M \hat{\theta}_{(\text{boot},m)}\,.\label{eq:theta_boot_avg}
\end{align}
$\hat{\theta}_\text{boot-avg}$ is analogous to $\hat{r}_\text{boot-avg}$ and $\hat{\mu}_\text{boot-avg}$ of \sref{sec:desc_of_orig_procedure} and it is the final estimate for $\theta$. Note that $\hat{\theta}_\text{boot-avg}$ is an unbiased estimator for $\theta$. 

The ensemble $E_\text{boot}$ is resampled with replacement $K$ times to get
\begin{align}
 E_{(\text{boot-boot},k)} &\equiv \left(\Big.\hat{\theta}_{(\text{boot-boot},k,1)},\dots,\hat{\theta}_{(\text{boot-boot},k,M)}\right)\,,\qquad\qquad\forall k\in\{1,\dots,K\}\,.
\end{align}
The elements of each of the double-bootstrapped ensembles are averaged to get
\begin{align}
 \hat{\theta}_{(\text{boot-boot-avg},k)} &\equiv \frac{1}{M}\sum_{m=1}^M \hat{\theta}_{(\text{boot-boot},k,m)}\,.\label{eq:theta_boot_boot_avg}
\end{align}
The sample standard deviation of $\hat{\theta}_{(\text{boot-boot-avg},k)}$-s is used as a candidate uncertainty estimate for $\hat{\theta}_\text{boot-avg}$:
\begin{align}
 \delta_\text{boot-boot} &\equiv \sqrt{\svar\left[\hat{\theta}_{(\text{boot-boot-avg},1)},\dots,\hat{\theta}_{(\text{boot-boot-avg},K)}\right]}\,,\label{eq:delta_boot_boot}
\end{align}
where $\svar$ represents the sample variance with the Bessel's correction. Roughly equivalently, based on the form of \eqref{eq:theta_boot_avg}, one can estimate the uncertainty in $\hat{\theta}_\text{boot-avg}$ as follows
\begin{align}
 \delta_\text{stderr-formula} &\equiv \sqrt{\frac{\svar\left[\hat{\theta}_{(\text{boot},1)},\dots,\hat{\theta}_{(\text{boot},M)}\right]}{M}}\,.\label{eq:delta_stderr_formula}
\end{align}
The claim of the present comment is that this uncertainty estimate is also flawed, in the same ways as $\delta_\text{boot-boot}$, since the $\hat{\theta}_{(\text{boot},m)}$-s are dependent on each other.

\paragraph{Theoretical results.} The following results are derived in \aref{appendix:derivations}:
\begin{subequations}\label{eq:box}
\begin{empheq}[box=\widefbox]{align}
 \var\left[\hat{\theta}_\text{boot-avg}\right] &= \frac{\sigma^2_X}{N} + \frac{N-1}{N}~\frac{\sigma^2_X}{NM} + \frac{\sigma^2_\epsilon}{M}\,,\label{eq:box_a}\\
 \E\left[\bigg.\var\left[\hat{\theta}_\text{boot-avg}~\Big|~D\right]\right] &= \frac{N-1}{N}~\frac{\sigma^2_X}{NM} + \frac{\sigma^2_\epsilon}{M}\,,\label{eq:box_b}\\
 \E\left[\delta^2_\text{boot-boot}\right] &= \frac{M-1}{M}\left[\frac{N-1}{N}~\frac{\sigma^2_X}{NM} + \frac{\sigma^2_\epsilon}{M}\right]\,,\label{eq:box_c}\\
 \E\left[\delta^2_\text{stderr-formula}\right] &= \frac{N-1}{N}~\frac{\sigma^2_X}{NM} + \frac{\sigma^2_\epsilon}{M}\,,\label{eq:box_d}
\end{empheq}
\end{subequations}
where $\E$ and $\var$ represent the population mean and population variance, respectively. In the right-hand-side of \eqref{eq:box_a}, roughly speaking, the first additive term captures the effect of finiteness of the original dataset $D$, the second term captures the randomness involved in resampling $D$ to produce the different bootstrapped datasets, and the third term captures the training errors.  It can be seen from \eqref{eq:box} that $\delta^2_\text{boot-boot}$ and $\delta^2_\text{stderr-formula}$ both miss the mc-stat uncertainty in the first term of \eqref{eq:box_a}, and only capture the uncertainties conditioned on $D$ in \eqref{eq:box_b}.\footnote{The extra factor of $(M-1)/M$ in \eqref{eq:box_c} represents a bias factor in the bootstrapping approach to estimating the variance of a sample mean. This can be corrected with a multiplicative factor of $M/(M-1)$, similar to Bessel's correction. Apply such a correction does not affect the conclusions drawn from the numerical experiment appreciably.}

Note that the factor of $1/M$ is present in the second and third terms of \eqref{eq:box_a} but absent in the first.
This is understandable, since one should not be able to suppress mc-stat just by increasing $M$. On the other hand, if one keeps $N$ fixed and increases $M$ indefinitely, then the double-bootstrapped error estimate will tend towards 0; this can be seen from \eqref{eq:box_c}.
This is true of the original non-toy example as well, for the double-bootstrapped estimate for the uncertainty in $\hat{r}_\text{boot-avg}(x)$ for any given $x$. As noted earlier, this is a sign that the double-bootstrapped error estimates cannot be capturing the mc-stat uncertainties (which originate from the finiteness of $N$) correctly.


In the original analyses in Ref.~\cite{ATLAS:2024jry,ATLAS:2024ynn}, having performed the double-bootstrapping approach for a given value of $M$, say $M_\mathrm{orig}$, one can reuse the trained neural networks to repeat the $\hat{r}_\text{boot-avg}$ estimation and the corresponding uncertainty quantification for other values of $M$ less than $M_\mathrm{orig}$. This way, one could empirically observe the $1/\sqrt{M}$ scaling of the double-bootstrapped error estimate for $\hat{r}_\text{boot-avg}$.\footnote{If the map from $\hat{r}_\text{boot-avg}$ to $\hat{\mu}_\text{boot-avg}$ has other sources of non-determinism, then the $1/\sqrt{M}$ scaling may not apply to the double-bootstrapped error estimate for $\hat{\mu}_\text{boot-avg}$.}

\paragraph{Numerical experiment.} The toy example was numerically simulated once with the following values for the various free parameters of the example:
\begin{align}
 \theta=5.0\,,\qquad \sigma_X = 100\,,\qquad \sigma_\epsilon = 0.01\,,\qquad N = 10^6\,,\qquad M=10^3\,,\qquad K=10^4\,.
\end{align}
The values of $\delta_\text{boot-boot}$ and $\delta_\text{stderr-formula}$ were $3.20\times 10^{-3}$ and $3.18\times 10^{-3}$, respectively (up to 3 significant figures). These numbers are in the same ballpark as $\sqrt{\E[\delta^2_\text{boot-boot}]}$ and $\sqrt{\E[\delta^2_\text{stderr-formula}]}$, computed, using \eqref{eq:box_c} and \eqref{eq:box_d}, to be approximately $3.18\times 10^{-3}$.

On the other hand, the true standard deviation of $\hat{\theta}_\text{boot-avg}$, as per \eqref{eq:box_a}, is approximately $1.00\times 10^{-1}$. Even if one does not trust \eqref{eq:box_a}, it is well known that, for the normal distribution, among all unbiased estimators of the mean $\theta$ based only on the dataset $D\equiv (X_1, \dots,X_N)$, the lowest variance is achieved by the sample mean of the $X_i$-s. The standard deviation of the sample mean is known to be $\sigma_X/\sqrt{N} = 0.1$, which is much larger than $\delta_\text{boot-boot}$ and $\delta_\text{stderr-formula}$.

The value of $\hat{\theta}_\text{boot-avg}$ was $4.96$ (up to 3 significant figures). If $\delta_\text{boot-boot}$ and $\delta_\text{stderr-formula}$ are assumed to be accurate estimates of uncertainty, then this represents a more than 10-sigma discrepancy from the true value of $\theta$. On the other hand, $\hat{\theta}_\text{boot-avg}$ is consistent with $\theta$ within its actual standard deviation of approximately $1.00\times 10^{-1}$. This numerical experiment (a) demonstrates that $\delta_\text{boot-boot}$ and $\delta_\text{stderr-formula}$ both significantly underestimate the uncertainty in $\hat{\theta}_\text{boot-avg}$ and (b) supports the claim that the double-bootstrapped uncertainty estimate misses the mc-stat uncertainty.

\section{Concluding remarks}
Note that if $\sigma^2_\epsilon/M >> \sigma_X^2/N$, then the double-bootstrapped uncertainty will approximately match the true uncertainty. More generally, the double-bootstrapped error estimation procedure in Ref.~\cite{ATLAS:2024ynn} would be acceptable if other sources of uncertainties dominate the mc-stat uncertainties and make them irrelevant. In Ref.~\cite{ATLAS:2024jry},  it is reported that mc-stat contributes only a small fraction of the overall uncertainty in the measurement. So, it is likely that (a) the mc-stat-contribution will continue to be small, and (b) the results of the analysis in Ref.~\cite{ATLAS:2024jry} will not be affected significantly, even after accounting for the point raised in the present comment. However, for precedent, even in such cases where mc-stat uncertainties can be ignored, it would be inaccurate to say that the mc-stat uncertainties are captured or estimated by the double-bootstrapping procedure.

While many applications of machine learning (ML) in high energy physics do not require the quantification of errors and uncertainties in the trained ML model, a few applications like NSBI do require uncertainty quantification. In some applications, like the one in Ref.~\cite{DES:2024xij}, the amount of simulated data available for (i) training, calibrating, and validating the ML models, and/or (ii) validating the inference pipeline (e.g., for constructing posteriors in a Bayesian context) is much larger than the amount of real experimental data on which the actual analysis is performed. One can often ignore mc-stat uncertainties in such situations. However, analyses in collider physics do not always fall under this category. Repetition- or ensemble-based techniques, possibly incorporating bootstrapping, are promising approaches to quantifying mc-stat and training uncertainties in ML. Given this context, Refs.~\cite{ATLAS:2024jry,ATLAS:2024ynn} represent an important milestone in the careful deployment of NSBI in collider physics.
It is important to note that the issue pointed out in the present comment is only a minor technicality, that could be fixed in a few different ways. 
One option would be to just use the standard bootstrapping technique and avoid merging multiple neural networks in the first place. However, one would be sacrificing the reduction of training errors achieved by such a merging (c.f. the $\sigma_\epsilon^2/M$ term in \eqref{eq:box_a}). Another option would be to train multiple networks on the original data and merge their outputs as before,\footnote{Here, all the networks can just be trained on the original dataset (possibly reshuffled each time), without any bootstrapping.} and repeat this procedure exactly in all the pseudo experiments performed using bootstrapped datasets. By choosing (a) the number of neural networks being merged and (b) the number of bootstraps for uncertainty quantification appropriately, one could potentially make this approach computationally feasible.\footnote{Even after such fixes are applied, the question of whether the surrogate distribution involved in the bootstrapping is sufficiently close to the corresponding true distribution (for the purposes of estimating the uncertainties in the ML-model) will still remain open, in the author's opinion; this is related to the intangible assumption discussed in \sref{sec:main_arguments}. Future investigations, possibly with realistic toy examples, will hopefully shed light on this matter.}

A third, possibly more ambitious approach to tackle mc-stat would be to treat the trained model for $r$, say $\hat{r}$, and other quantities computed using $\hat{r}$ (e.g., the approximate likelihood function that maps an individual event to a function of parameters) simply as useful learned transformations or summaries of the data, without an inherent associated uncertainty. The subsequent analyses could be performed by comparing the observed data against simulations, via such summaries. Various systematic uncertainties, including mc-stat, could potentially be incorporated in the simulation-based estimation of the theoretical distributions of such summaries.\footnote{The simulated data used for such estimations will have to be independent of the data used to train $\hat{r}$.} However, this approach might require one to flesh out the details of performing statistical analysis using function-summaries of events, which could be non-trivial.

\section*{Code and data availability}
The code used to perform the numerical experiment is available at the following URL: \url{https://gitlab.com/prasanthcakewalk/double-bootstrapping-toy-example}.

\section*{Acknowledgments}
The author thanks Oz Amram, Stephen Mrenna, Kevin Pedro, Gabriel Perdue, and in particular, Nicholas Smith and Manuel Szewc for useful discussions and/or feedback on this comment. The following open source software and tools were used directly in performing the numerical experiment in this comment: Python \cite{Python}, NumPy \cite{NumPy}.

This document was prepared using the resources of the Fermi National Accelerator Laboratory (Fermilab), a U.S. Department of Energy, Office of Science, Office of High Energy Physics HEP User Facility. Fermilab is managed by Fermi Forward Discovery Group, LLC, acting under Contract No. 89243024CSC000002. The author is supported by the U.S. Department of Energy, Office of Science, Office of High Energy Physics QuantISED program under the grants ``HEP Machine Learning and Optimization Go Quantum'', Award Number 0000240323, and ``DOE QuantiSED Consortium QCCFP-QMLQCF'', Award Number DE-SC0019219.

%


\appendix
\section{Derivations for the toy example}\label{appendix:derivations}

As a preface, Bienaym\'{e}'s identities, given by
\begin{align}
 \var\left[\sum_{i=1}^n Y_i\right] &= \sum_{i=1}^n \var[Y_i] + \sum_{i\neq j} \cov[Y_i~,~Y_j]\,,\\
 \cov\left[\sum_{i=1}^n Y_i~,~\sum_{j=1}^m Z_j\right] &= \sum_{i=1}^n\sum_{j=1}^m \cov[Y_i~,~Z_j]\,,
\end{align}
are used several times in the following derivations. Here $\cov$ represents population covariance. Likewise, the laws of total variance and total covariance, given by
\begin{align}
 \var\left[Y\right] &\equiv \E\left[\big.\var[Y~|~Z]\right] + \var\left[\big.\E[Y~|~Z]\right]\qquad\qquad\qquad\qquad\qquad\text{and}\\
 \cov\left[Y_1, Y_2\right] &\equiv \E\left[\big.\cov[Y_1, Y_2~|~Z]\right] + \cov\left[\big.\E[Y_1~|~Z]~,~\E[Y_2~|~Z]\right]\,,
\end{align}
respectively, are also used extensively.

From \eqref{eq:theta_boot}, it can be seen that for $m\neq m'$,
\begin{align}
 &\cov\left[\Big.\hat{\theta}_{(\text{boot},m)}~,~\hat{\theta}_{(\text{boot},m')}\right] = \frac{1}{N^2}\sum_{i,i'=1}^N \cov\left[X_{(\text{boot},m,i)}~,~X_{(\text{boot},m',i')}\right]\\
 &= \cov\left[X_{(\text{boot},m,i)}~,~X_{(\text{boot},m',i')}\right]\\
 &= \underbrace{\E\left[\bigg.\cov\left[X_{(\text{boot},m,i)}~,~X_{(\text{boot},m',i')}~\Big|~D\right]\right]}_{=\,0} + \cov\left[\bigg.\E\left[X_{(\text{boot},m,i)}~\Big|~D\right]~,~\E\left[X_{(\text{boot},m',i')}~\Big|~D\right]\right]\\
 &= \var\left[\Big.\smean[X_1,\dots,X_N]\right] = \frac{\sigma^2_X}{N}\,, \label{eq:cov_boot_m}
\end{align}
where $\smean$ represents the sample mean. Similarly, from \eqref{eq:theta_boot}, it can be seen that
\begin{align}
 \var\left[\Big.\hat{\theta}_{(\text{boot},m)}\right] &= \frac{1}{N}\var[X_{(\text{boot},m,i)}] + \frac{N-1}{N}\cov\left[X_{(\text{boot},m,i)}~,~X_{(\text{boot},m,i')}\right]\Bigg|_{i'\neq i} + \sigma^2_\epsilon\\
 &= \frac{\sigma^2_X}{N} + \frac{N-1}{N}~\frac{\sigma^2_X}{N} +\sigma^2_\epsilon\,. \label{eq:var_boot_m}
\end{align}
Now, from \eqref{eq:theta_boot_avg} it can be seen that
\begin{align}
 \var\left[\hat{\theta}_\text{boot-avg}\right] &\equiv \frac{1}{M}\var\left[\Big.\hat{\theta}_{(\text{boot},m)}\right] + \frac{M-1}{M}\cov\left[\Big.\hat{\theta}_{(\text{boot},m)}~,~\hat{\theta}_{(\text{boot},m')}\right]\bigg|_{m\neq m'}\,.
\end{align}
Plugging \eqref{eq:cov_boot_m} and \eqref{eq:var_boot_m} into the previous equation leads to \eqref{eq:box_a}. From the law of total variance,
\begin{align}
 \E\left[\bigg.\var\left[\hat{\theta}_\text{boot-avg}~\Big|~D\right]\right] &= \var\left[\hat{\theta}_\text{boot-avg}\right] - \var\left[\bigg.\E\left[\hat{\theta}_\text{boot-avg}~\Big|~D\right]\right]\\
 &= \var\left[\hat{\theta}_\text{boot-avg}\right] - \var\left[\Big.\smean[X_1,\dots,X_N]\right] = \var\left[\hat{\theta}_\text{boot-avg}\right] - \frac{\sigma_X^2}{N}\,.
\end{align}
Plugging \eqref{eq:box_a} into the previous equation leads to \eqref{eq:box_b}.

Next, from \eqref{eq:theta_boot_boot_avg}, it can be seen that for $k\neq k'$
\begin{align}
 &\cov\left[\Big.\hat{\theta}_{(\text{boot-boot-avg},k)}~,~\hat{\theta}_{(\text{boot-boot-avg},k')}\right] = \frac{1}{M^2}\sum_{m, m'=1}^M \cov\left[\hat{\theta}_{(\text{boot-boot},k,m)}~,~\hat{\theta}_{(\text{boot-boot},k',m')}\right]\\
 &= \cov\left[\hat{\theta}_{(\text{boot-boot},k,m)}~,~\hat{\theta}_{(\text{boot-boot},k',m')}\right]\\
 \begin{split}
 &= \underbrace{\E\left[\bigg.\cov\left[\hat{\theta}_{(\text{boot-boot},k,m)}~,~\hat{\theta}_{(\text{boot-boot},k',m')}~~\Big|~~E_\mathrm{boot}\right]\right]}_{=\,0}\\
 &\qquad\qquad\qquad\qquad\qquad+\cov\left[\bigg.\E\left[\hat{\theta}_{(\text{boot-boot},k,m)}~\Big|~E_\mathrm{boot}\right]~,~\E\left[\hat{\theta}_{(\text{boot-boot},k',m')}~\Big|~E_\mathrm{boot}\right]\right]
 \end{split}\\
 &= \var\left[\Big.\text{sample mean of $\left(\Big.\hat{\theta}_{(\text{boot},1)}, \dots,\hat{\theta}_{(\text{boot},M)}\right)$}\right] = \var\left[\hat{\theta}_\text{boot-avg}\right]\,.\label{eq:cov-theta-bootboot}
\end{align}
Furthermore, from \eqref{eq:theta_boot_boot_avg}, it can be seen that
\begin{align}
\begin{split}
 \var\left[\Big.\hat{\theta}_{(\text{boot-boot-avg},k)}\right] &= \frac{1}{M}\var\left[\hat{\theta}_{(\text{boot-boot},k,m)}\right] \\
 &\qquad\qquad\quad+ \frac{M-1}{M}\,\cov\left[\hat{\theta}_{(\text{boot-boot},k,m)}~,~\hat{\theta}_{(\text{boot-boot},k,m')}\right]\bigg|_{m\neq m'}
\end{split}\\
 &= \frac{\var\left[\hat{\theta}_{(\text{boot},m)}\right]}{M} + \frac{M-1}{M}\,\var\left[\hat{\theta}_\text{boot-avg}\right]\,.\label{eq:var-theta-bootboot}
\end{align}
One can rewrite $\delta^2_\text{boot-boot}$ in \eqref{eq:delta_boot_boot} as
\begin{align}
 \delta^2_\text{boot-boot} &=\frac{1}{K(K-1)}\sum_{k < k'} \left(\hat{\theta}_{(\text{boot-boot-avg},k)} - \hat{\theta}_{(\text{boot-boot-avg},k')}\right)^2\,.
\end{align}
This, with \eqref{eq:cov-theta-bootboot} and \eqref{eq:var-theta-bootboot}, leads to
\begin{align}
 \E\left[\delta^2_\text{boot-boot}\right] &= \var\left[\hat{\theta}_{(\text{boot-boot-avg},k)}\right] - \cov\left[\hat{\theta}_{(\text{boot-boot-avg},k)}~,~\hat{\theta}_{(\text{boot-boot-avg},k')}\right]\bigg|_{k\neq k'}\\
 &= \frac{1}{M}\left(\bigg.\var\left[\hat{\theta}_{(\text{boot},m)}\right] - \var\left[\hat{\theta}_\text{boot-avg}\right]\right)\,.
\end{align}
Plugging \eqref{eq:var_boot_m} and \eqref{eq:box_a} into the previous equation leads to \eqref{eq:box_c}.
Likewise, for the error estimate in \eqref{eq:delta_stderr_formula},
\begin{align}
 \E\left[\delta^2_\text{stderr-formula}\right] &= \frac{1}{M}\,\left(\var\left[\hat{\theta}_{(\text{boot},m)}\right] - \cov\left[\hat{\theta}_{(\text{boot},m)}~,~\hat{\theta}_{(\text{boot},m')}\right]\bigg|_{m\neq m'}\right)\,.
\end{align}
Plugging \eqref{eq:cov_boot_m} and \eqref{eq:var_boot_m} into the previous equation leads to \eqref{eq:box_d}.


\begin{thebibliography}{9}
\bibitem{ATLAS:2024jry}
ATLAS Collaboration,
``Measurement of off-shell Higgs boson production in the $H^*\rightarrow ZZ\rightarrow 4\ell$ decay channel using a neural simulation-based inference technique in 13\,TeV $pp$ collisions with the ATLAS detector,''
Rept. Prog. Phys. \textbf{88}, no.5, 057803 (2025)
doi:10.1088/1361-6633/adcd9a
[\href{https://arxiv.org/abs/2412.01548v2}{arXiv:2412.01548v2 [hep-ex]}].

\bibitem{ATLAS:2024ynn}
ATLAS Collaboration,
``An implementation of neural simulation-based inference for parameter estimation in ATLAS,''
\href{https://arxiv.org/abs/2412.01600v1}{arXiv:2412.01600v1 [hep-ex]}.

\bibitem{ATLAS:2020ocz}
 ATLAS Collaboration,
``Recommendations for the Modeling of Smooth Backgrounds,''
ATL-PHYS-PUB-2020-028,
\url{https://cds.cern.ch/record/2743717/}.

\bibitem{DES:2024xij}
N.~Jeffrey \textit{et al.} [DES],
``Dark Energy Survey Year 3 results: likelihood-free, simulation-based $w$CDM inference with neural compression of weak-lensing map statistics,''
Mon. Not. Roy. Astron. Soc. \textbf{536}, no.2, 1303-1322 (2025)
doi:10.1093/mnras/stae2629
[\href{https://arxiv.org/abs/2403.02314}{arXiv:2403.02314 [astro-ph.CO]}].

\bibitem{Python}
G.~Van~Rossum and F.~L.~Drake Jr,
``Python reference manual,''
Centrum voor Wiskunde en Informatica Amsterdam (1995).

\bibitem{NumPy}
C.~R.~Harris, K.~J.~Millman, S.~J.~van~der~Walt, R.~Gommers, P.~Virtanen, D.~Cournapeau, E.~Wieser, J.~Taylor, S.~Berg, N.~J.~Smith, R.~Kern, M.~Picus \textit{et al.},
``Array programming with NumPy,''
Nature {\bf 585} (7825), 357 (2020)
doi:10.1038/s41586-020-2649-2
[\href{https://arxiv.org/abs/2006.10256}{arXiv:2006.10256 [cs.MS]}].

\end{thebibliography}
\end{document}